\documentclass[]{pnas-new}

\templatetype{pnasresearcharticle}

\title{Leveraging Artificial Intelligence for Democratic Discourse: Chat Interventions can Improve Online Political Conversations at Scale}

\author[a]{Lisa P. Argyle}
\author[b]{Chris Bail}
\author[a]{Ethan C. Busby}
\author[a]{Joshua R. Gubler}
\author[c]{Thomas Howe}
\author[c]{Christopher Rytting}
\author[d]{Taylor Sorensen}
\author[c]{David Wingate}

\affil[a]{Department of Political Science, Brigham Young University}
\affil[b]{Department of Sociology, Political Science, and Public Policy, Duke University}
\affil[c]{Department of Computer Science, Brigham Young University}
\affil[d]{Department of Computer Science, University of Washington}

\leadauthor{Argyle} 


\begin{abstract}
Abstract:  
Political discourse is the soul of democracy, but misunderstanding and conflict can fester in divisive conversations. The widespread shift to online discourse exacerbates many of these problems and corrodes the capacity of diverse societies to cooperate in solving social problems. Scholars and civil society groups promote interventions that make conversations less divisive or more productive, but scaling these efforts to online discourse is challenging.  We conduct a large-scale experiment that demonstrates how online conversations about divisive topics can be improved with artificial intelligence tools. Specifically, we employ a large language model to make real-time, evidence-based recommendations intended to improve participants' perception of feeling understood. These interventions improve reported conversation quality, promote  democratic reciprocity, and improve the tone, without systematically changing the content of the conversation or moving people's policy attitudes. 

\end{abstract}

\dates{This manuscript was compiled on \today}
\doi{\url{www.pnas.org/cgi/doi/10.1073/pnas.XXXXXXXXXX}}

\begin{document}

\maketitle
\thispagestyle{firststyle}
\ifthenelse{\boolean{shortarticle}}{\ifthenelse{\boolean{singlecolumn}}{\abscontentformatted}{\abscontent}}{}

\dropcap{S}ocial scientists have long observed that ``conversation is the soul of democracy'' \cite{doi:10.1177/2057047316628310,deweypublicproblems}. Interpersonal  discussions across social divides can help diverse groups of people peacefully identify solutions to shared problems, avoid violent conflict, and come to understand one another better \cite{deweypublicproblems,muddiman2017personal, gutmann2004deliberative,sydnor2019disrespectful, finkel_political_2020, mutz2002cross, jacobs2009talking}. Historically, these conversations have occurred face-to-face \cite{jacobs2009talking}, but online conversations now play a central role in public dialogue. More than 100 billion messages are sent every day on Facebook and Instagram alone \cite{mosseri2020}, and approximately 7 billion conversations occur daily on Facebook Messenger \cite{bleu2023}. Such conversations can have far-reaching impact. Some of the largest social movements in human history have emerged out of sprawling conversations on social media, and discussions between high profile social media users can shape the stock market, politics, and many other aspects of human experience \cite{harlow2011,mundt2018,jiao2020,kessel2020}. The internet thus has the capacity to empower an ever-increasing number of people to communicate and deliberate together.

However, there is growing concern that much of online conversation does the opposite \cite{garsd2019,settle2018,bail2021,sun2021}. Nearly half of social media users report observing mean or cruel behavior, and many indicate that online divisiveness and incivility complicate a variety of relationships in their lives -- with family, friends, and work colleagues \cite{rainie2012}. As such, many members of the public either avoid online discussions about politics or unwittingly find themselves arguing online in a corrosive, unconstructive manner \cite{millerconover2015,lee2021, santoro2022promise, carlson2022goes}. Such rhetoric has been linked to partisan violence \cite{Kalmoe2018,kalmoe2022radical}, disengagement from politics and public life \cite{krupnikov2022other,carlson2022goes}, and reduced capacity to find compromise \cite{wolf2012incivility}.

These online political conversations are a far cry from the types of conversations scholars identify as the foundation of deliberative democracy \cite{mendelberg_2002,gutmann2004deliberative,scudder2020beyond,dobson2014listening}. Democratic deliberation demands conversations built on what deliberative scholars call ``democratic reciprocity'': a willingness to grant political opponents the same right to express and advocate their views in the public sphere that we hope they will grant us \cite{gutmann2004deliberative}. This does not imply conversations that end with agreement, or condoning ideas that are problematic, but instead suggests conversational willingness to listen to and engage in good faith with with those who have political opinions that differ from our own. 

Thankfully, scholarship on how to facilitate these types of conversations continues to grow across disciplines \cite{mutz2002cross,broockman2016durably,levenduskystecula2021,wojcieszak2020,amsalem2022,reis2017toward,ruan2020can,itzchakov2022foster,livingstone2020they}. These studies identify a range of strategies to increase the likelihood that members of rival groups thoughtfully listen to and engage with others' perspectives. Though conversations built on such strategies rarely  result in immediate resolution of political problems and disagreements, many scholars see them as a necessary condition for increasing mutual understanding, compromise, and coalition-building. That is, ``hearing the other side'' \cite{mutz2006hearing} can be democratically beneficial even if disagreement remains, providing a variety of broader benefits related to social cohesion and democracy \cite{mutz2002cross, jacobs2009talking, santoro2022promise,levenduskystecula2021}. 

In what follows, we present the results of a field experiment that employed cutting-edge artificial intelligence tools -- in this case, the large language model (LM) GPT-3 -- to \textit{scale up} evidence-based conversation-improving interventions. We invited proponents and opponents of gun regulation in the United States into online conversations, randomly assigning a pre-trained chat assistant powered by GPT-3 to some of the participants. We show that intervention by the chat assistant, which recommended real-time, context-aware, and evidence-based ways to rephrase messages, improved democratic reciprocity in conversation, particularly for the partner of the person assigned the AI assistant. 

\subsection*{AI Tools in Social Science}

Political actors and social scientists increasingly use artificial intelligence tools to influence and study the social world \cite{tappin_wittenberg_hewitt_berinsky_rand_2022,aggarwal20232,munger2017tweetment,bail2018exposure}. Language models like ChatGPT and others highlight the ability of artificial intelligence to generate human-sounding text and perform tasks previously thought impossible \cite{tiku2022}. Given their potential to identify and replicate complex patterns in text, LMs provide a promising new way to explore social outcomes \cite{argyle22}. One important advance of these models is their capacity for ``few-shot'' learning, or their ability to learn to perform a task from just a few exemplars without requiring parameter updates \cite{brown2020language}. 

While many observers are rightfully concerned about the negative effects of biases present in LMs and other AI tools \cite{bender2021dangers,kleinberg_human_2018,panchArtificialIntelligenceAlgorithmic2019,caliskan2017semantics,obermeyerDissectingRacialBias2019}, the same model features that generate these biases also enable LMs to produce text that is nuanced and multifaceted in its representation of a range of people, tones, ideas, and attitudes \cite{argyle22}. Prior AI-in-the-loop applications have demonstrated that AI can help people be more empathetic in peer mental health support conversations \cite{sharma2023human}, and that AI-induced reflection and restatement can improve the quality of conversations \cite{kriplean2012, kim2021, YEOMANS2020}. We build on that work, as well as on frameworks developed by scholars like Fishkin et. al. \cite{fishkin2019deliberative}, to demonstrate that dynamic, real-time, and context-aware LM-generated recommendations can improve the quality of political conversations by helping people communicate their willingness to respect, acknowledge, and be open to the views of their political opponents. 

\subsection*{Strategies to Improve the Perception of Being Understood and Democratic Reciprocity}

Listening is a core---if understudied---element of democratic politics \cite{dobson2014listening, scudder2020beyond}. It can make the policy process more efficient and empower those who feel marginalized from the policy process \cite{esaiasson2017responsiveness}. Understanding and acknowledging others' perspectives -- and feeling understood and acknowledged oneself -- has deep connections to many approaches to conflict resolution and deliberative democracy \cite{itzchakov2022foster, livingstone2020they,mendelberg_2002}. For some scholars, a commitment to this type of listening and acknowledgement is a necessary first step to democracy, without which productive and constructive forms of political decision-making cannot exist \cite{mendelberg_2002, gutmann2004deliberative}. 

In this research project, we use an LM in political conversations among policy opponents to increase people's perception that they have been listened to and understood by their conversation partner. Although all conversations across lines of difference do not reduce conflict and divisiveness \cite{paluck2019contact}, the feeling of being understood has been shown to generate a host of positive social outcomes \cite{reis2017toward,gordon2016,livingstone2020they,pollmann2009, minson2022}. 
Research suggests a number of specific, actionable conversation techniques to effectively increase the perception of being understood \cite{itzchakov2022foster,ruan2020can,reis2017toward, YEOMANS2020}, used in a variety of settings worldwide \cite{hartman2022interventions}. As we discuss in further detail below, these include strategies like increasing general politeness, validating the legitimacy of others to have different views, and simply restating another person's position to signal that they have been correctly heard and understood \cite{itzchakov2022foster,ruan2020can,reis2017toward}. 

In practice, civil society and academic approaches to improving conversations typically require trained moderators and instructors to teach, model, and help people develop and practice these skills. While effective, such interventions reach only a tiny fraction of those caught in divisive conversations daily. The challenge is implementation at scale: helping individuals recognize and remember how to apply these techniques, and/or find the will to apply them, in the moment of a (heated) real-world conversation. Additionally, research shows that the benefits of such conversations \textit{do not require persuasion} or agreement between participants on the issues discussed. Accordingly, our use of AI in this experiment does not seek to change participants' minds; we suggest this as a model for how AI can be employed without pushing a particular political or social agenda. Defining high-quality conversations as those in which people feel like they have been respected and understood by their discussion partner is an intentional response to well-founded concerns about normative goals that over-emphasize  civility or prioritize ideologically-motivated persuasion as a means of depolarization. \cite{sydnor2019disrespectful, krupnikov2022other, broockmankalla2022, kreiss2023polarization}.

Importantly, communicating openness and respectful listening are outcomes that we expect to have an impact beyond the context of a single brief conversation with one other person. We expect the perception of feeling understood to also promote a sense of \textit{democratic reciprocity} which ``asks citizens [...] to try to justify their views to one another and to treat with respect those who make a good-faith effort to engage in this mutual enterprise, even when they cannot resolve their disagreements'' \cite{gutmann2004deliberative}. In its strongest form, reciprocity entails an expectation that citizens make a sincere effort to engage with the public reasons provided by their opponents, such that they are open to the possibility of being persuaded by or compromising with those views. Theorists emphasize benefits to democracy generally when individuals make good-faith efforts to understand others' views -- even when they cannot resolve disagreements and remain sharply divided on political topics \cite{gutmann2004deliberative, mutz2006hearing, scudder2020beyond, mendelberg_2002}. In this sense, increasing democratic reciprocity can be seen as an important precursor to the reduction of partisan animosity or polarization; as citizens begin to acknowledge and express respect for the democratic legitimacy of their opponents in the political system, they may be more likely to support democratic institutions as legitimate processes to resolve conflict and reach compromise.

 \section*{Hypotheses} 

We developed an AI chat assistant to act as an at-scale, real-time moderator in divisive political conversations. The assistant makes tailored suggestions on how to rephrase specific texts in the course of a live, online conversation, without fundamentally affecting the policy content or position taken in the messages. 
The suggestions are based on three specific techniques from the literatures on listening, understanding, and deliberative democracy mentioned earlier: \textbf{restatement}, simply repeating back a person's main point to demonstrate understanding; \textbf{validation}, affirming the legitimacy of others holding different opinions without requiring explicit statements of agreement (e.g. ``I can see you care a lot about this issue''); and \textbf{politeness}, modifying the statement to use more polite language. 

Our \href{https://osf.io/k8hnr/?view_only=b2323e1efb2047febebf33dd3f6fe421}{pre-registered expectations} are that individuals in chats with political opponents where one participant has the rephrasing assistance of our AI tool will report feeling more understood themselves (higher conversation quality) and acknowledge the perspectives of others more readily (increased democratic reciprocity), even if they still disagree with their chat partner, than those in untreated chats. We expect no treatment effect on policy attitudes.

\begin{figure*}
\centering
\includegraphics[width = \linewidth]{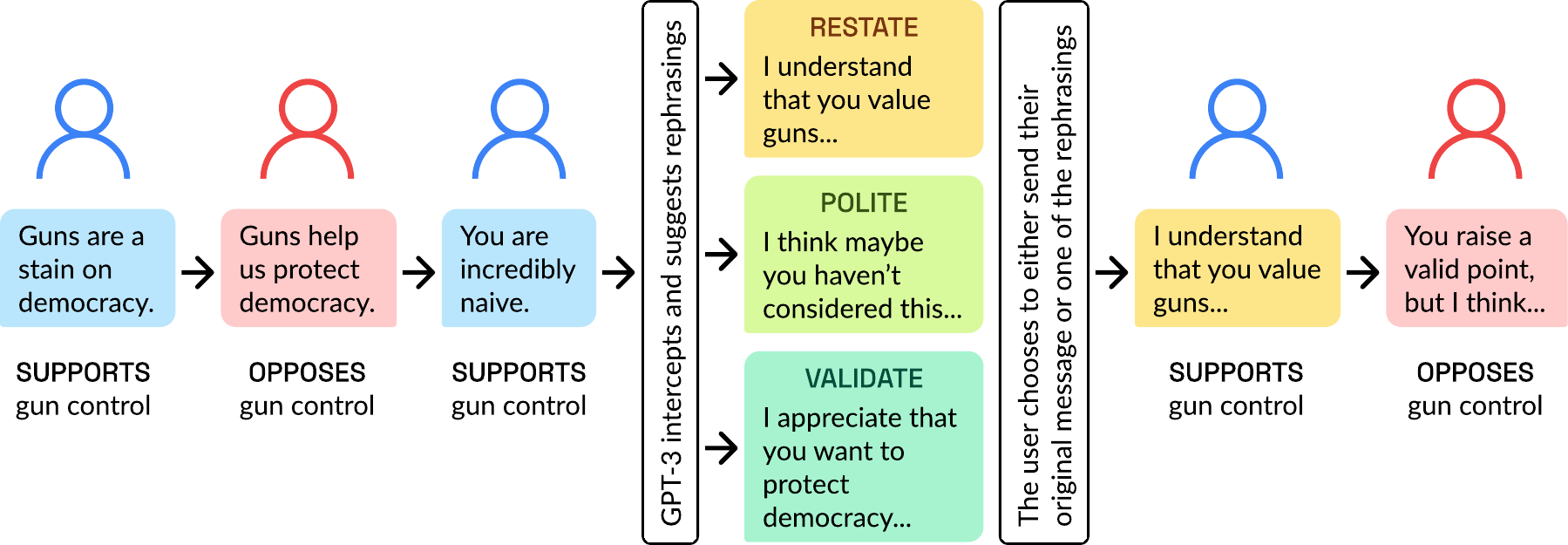}
\caption{\textbf{Treated Conversation Flow}: Respondents write messages unimpeded until one partner receives a rephrasing prompt for the first message longer than four words, and every other conversational turn thereafter. The chat assistant intercepts the treated user's message, using GPT-3 to propose evidence-based alternative phrasings, while retaining the semantic content. It suggests three randomly ordered alternatives to the author of the message and presents the opportunity to accept or edit any of these rephrasing suggestions or send their original message. Their choice is sent to their partner and the conversation continues.}
\label{fig:flow}
\end{figure*}

\section*{Study Design}

We test these hypotheses in an online chat experiment about gun regulation in the United States. We asked participants to discuss gun policy because it is a divisive issue that has near constant salience to American political debates \cite{obrien2013,lacombe2019,filindra2021,lacombe2021}. After a brief pre-survey, participants were matched with another respondent with whom they disagreed about gun policies. 

Once matched, conversation pairs were randomly assigned to the treatment or control condition, and partners proceeded to have a conversation. In a treated conversation, one participant received intermittent suggestions of ways to rephrase their message prior to sending it to their partner. Figure \ref{fig:flow} shows how the rephrasing prompts from GPT-3 fit into the conversational flow. Participants could choose to send one of three AI-suggested alternatives, their original message, or edit any message. 
 
After completing the chat, respondents were routed to another survey that measured their impressions of conversation quality, levels of democratic reciprocity towards those who disagree with them on gun regulation, and the same measures of their views of gun regulation as in the pre-chat survey. By conversational quality, we mean to measure participants' perceptions of feeling understood by their partner and the respectfulness of the conversation they just had. We use five items that ask participants to rate the degree to which they ``felt heard and understood by my partner'' and other related questions. 

To measure democratic reciprocity, we asked another four questions designed to create an index of participants' willingness to respect the views of their opponents in the broader political system. As opposed to the conversational quality items, which focus on the individual's experience with a specific other person in a defined conversation, these focus on attitudes towards their policy opponents generally. The survey items include evaluations of agreement with statements like ``I respect the opinions of people who disagree with me on gun regulation'' and ``It is important to understand people who disagree with me on gun regulation by imagining how things look from their perspective.'' Both scales have good psychometric properties (see the Supporting Information for more details and results by each question).

\begin{figure*}
\begin{center}
\includegraphics[width=.7\linewidth]{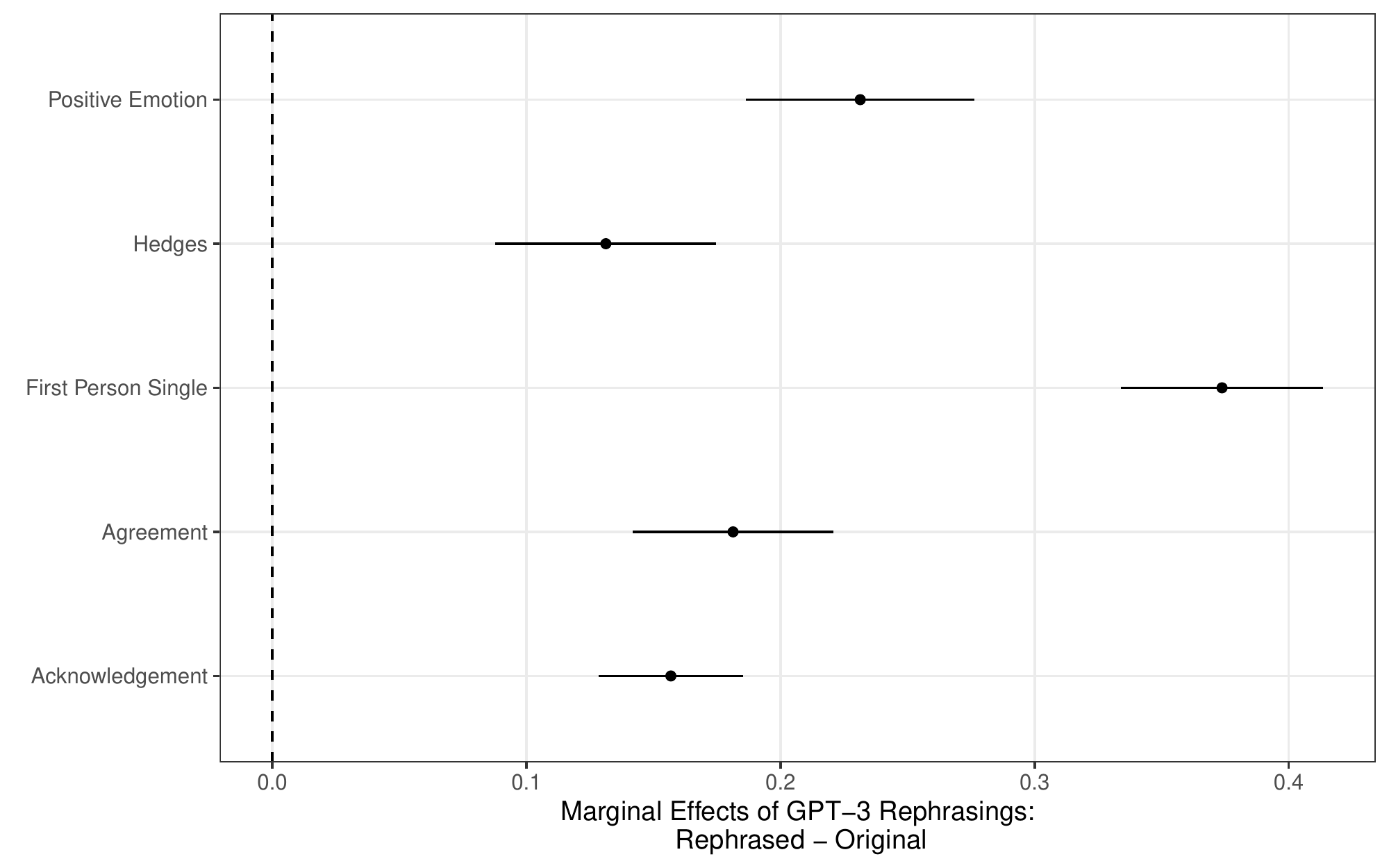}
\caption{\textbf{Text analysis of rephrased messages tone}: Marginal difference, with 95\% confidence intervals, between rephrased message scores on five \texttt{politeness} package features, and baseline scores from participants' original messages they would have sent had they not chosen the rephrasing.}
\label{fig:rephrasings}
\end{center}
\end{figure*}

\section*{Results}

In October 2022, 1,574 people participated in our field experiment. On average, 12 total messages were sent in each conversation with a total of 2,742 AI rephrasings suggested. AI-suggested rephrasings were accepted by chat participants two-thirds (1,798) of the time. Accepted rephrasings were roughly evenly split between the restate (30\%), validate (30\%), and politeness (40\%) interventions.

\subsection*{AI Rephrasings: Tone and Topic}

We first analyze the text of these conversations to verify that the chat assistant functioned as intended. In particular, we explore 1) the degree to which AI-rephrased messages chosen by participants differ from  messages they would have sent otherwise in their \textbf{tone}, and 2) the degree to which these messages differ in terms of \textbf{topic}. If the assistant worked as intended, then the rephrasings should be more polite and validating than the original messages, but no different in topic.

To explore differences in tone, we identified all 899 original messages replaced by a rephrasing and used the \texttt{politeness} package in \texttt{R} \cite{yeomans2018} to generate scores for both original and the AI-rephrased messages selected by users to replace their original text. Based on recent research by Yeomans et al. \cite{YEOMANS2020}, we generated scores for five text features, expecting rephrased messages to score higher on average on each feature than original messages: positive emotion, hedges, first person singular, agreement, and acknowledgement. We then estimated simple OLS models, regressing a binary feature score for each feature on a binary variable indicating whether the message was rephrased or original. Figure \ref{fig:rephrasings} presents the marginal average difference between rephrased messages and the original text on each of these text feature outcomes. AI-rephrased messages chosen by participants contained more of each of these features than the original messages participants would have otherwise sent.

\begin{figure*}[ht]
\centering
\includegraphics[width = 0.97\linewidth]{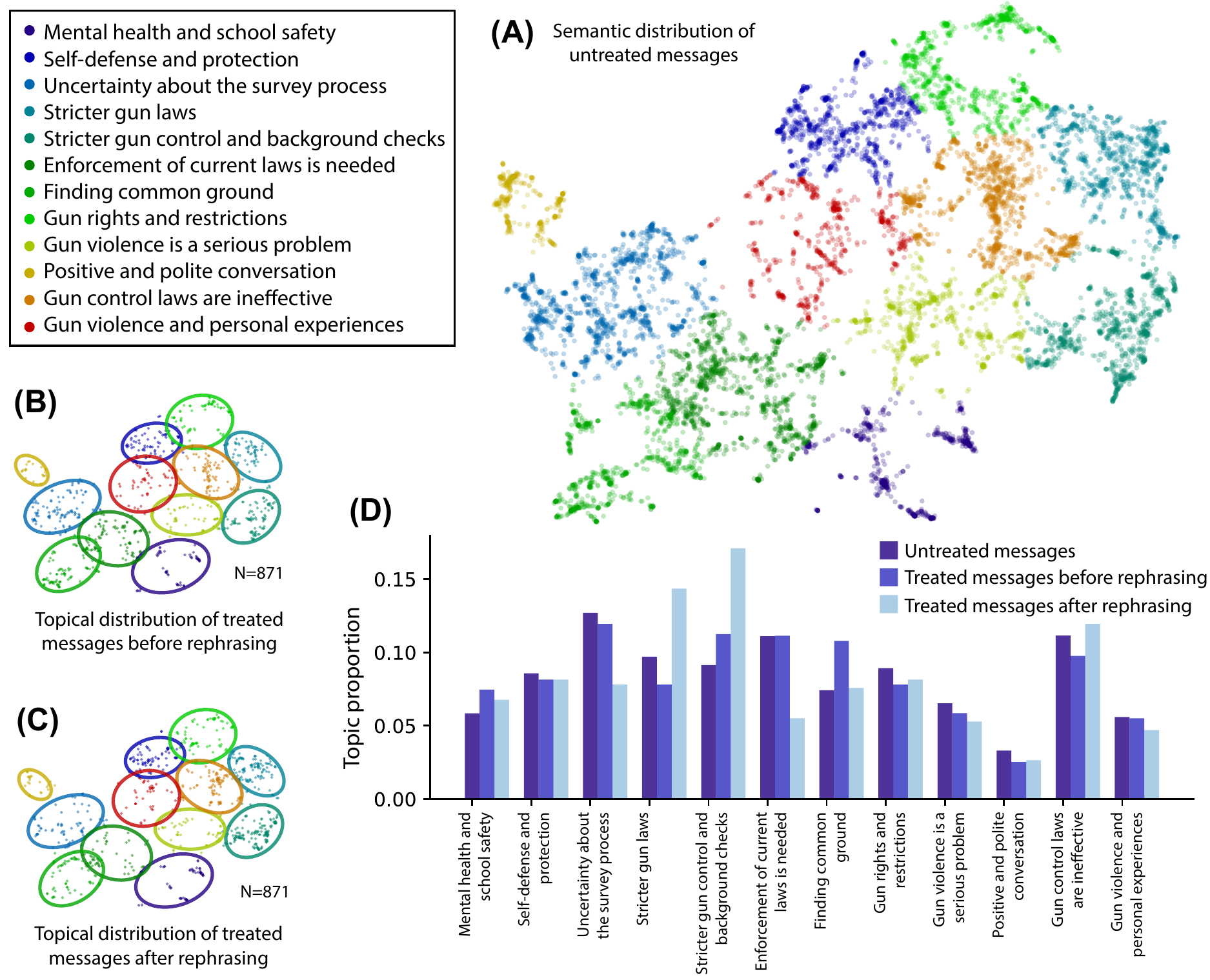}
\caption{\textbf{Analysis of Semantic Content of Messages}. \textbf{Panel A} presents a visualization of the topical distribution of messages sent on the platform. Each point is the semantic embedding of a message; points that are close to each other represent messages that are semantically similar. Messages are clustered with k-means, and clusters are automatically labeled by GPT-4; see SI for technical details. As demonstrated by the figure, the conversations spanned a wide range of sub-topics about gun control, including background checks, school safety, the role of guns in school, mental health, and enforcement issues surrounding gun ownership.  Additional topic clusters show general conversational dynamics, such as introductory or closing material.
\textbf{Panel B} and \textbf{Panel C} graphically show the distribution of messages selected for treatment and the distribution of the corresponding rephrased messages. Both sets of messages are similarly distributed, both to each other and to the untreated messages shown in Panel A.
\textbf{Panel D} quantitatively shows the topic proportions; statistical analysis shows that the distributions are not significantly different.}
\label{fig:umap}
\end{figure*}

To confirm that AI rephrasings changed message tone \textit{but not topic}, we used an automated pipeline and a variety of ML techniques discussed further in the Supporting Information to cluster all messages sent with more than 4 words by topic in a 2D space. We then used GPT-4 to automatically generate a short summary of the content of each cluster. Panel (A) of Figure~\ref{fig:umap} shows the topic clusters and corresponding GPT-4-generated labels. The labels show that the vast majority of messages sent on the platform were on-topic; additional manual checking confirmed this. Panels (B) and (C) show the distribution of treated messages before and after rephrasing; as can be seen, messages that were (randomly) selected for treatment, and their corresponding AI-rephrasings, are spread evenly throughout the semantic space. This indicates that rewritten messages did not fundamentally alter topical distribution, nor were there obvious degeneracies (such as mapping all rewritten points to a single cluster, or creating fundamentally new clusters).  Panel (D) shows the quantitative topic proportions of all three types of messages; the distributions are not significantly different ($\chi^2=0.150$, N=871, $p$=1.00).

\subsection*{Treatment Effects}

We now turn to our main results: a presentation of the effects of the rephrasings on both \textbf{conversational quality} (the degree to which individuals felt they were understood and respected in the conversation) and \textbf{democratic reciprocity} (the degree to which they were willing to grant this same listening and respect to political opponents).

Recall that random assignment in our experiment occurred at the conversation level, generating treated and control chats. However, only one person in a ``treated'' conversation was assigned the chat assistant intervention. Therefore, we present results for three effects: one for the person who used the assistant (``GPT-3 Self''), one for those whose partner used the assistant (``GPT-3 Partner''), and another for those in control conversations with no assistant (``Control'').

By design, conversations were expected to continue until the treated individual in the chat received four rephrasing prompts; equivalently, control conversations were set to finish after one partner would have received four interventions, had they been provided. However, in practice, only 698 (44\%) of participants were in chats that lasted the full intended length (see the Supplemental Materials for further discussion). Thus, as is common in field experiments like ours, some participants assigned to be treated (to participate in chats lasting at least four AI-suggested rephrasings long) only received partial treatment: they or their partner left the chat platform early for a variety of reasons. Nearly all of the participants whose conversations ended early still completed the post-survey. 

As all participants were in conversations in which early departure was equally possible, we follow Gerber and Green \cite{gerber2012field} in calculating placebo-controlled treatment effects for separate subgroups of the study population based on the number of interventions they received if they were in a treated chat, or would have received had they not been in a control conversation. As such, for each subgroup, we calculate simple means and confidence intervals for both ``GPT-Self'' and ``GPT-Partner'' participants in the treated chats, and contrast them to those in the control chats of equivalent lengths. As Gerber and Green \cite{gerber2012field} note, 
these mean differences are causally identified treatment effects within each subgroup of the data under the assumption that the treatment -- rephrasing interventions -- is unrelated to people's persistence in the conversation. See the Supporting Information for several tests supporting this assumption.

\begin{figure*}[ht]
\includegraphics[width=1\linewidth]{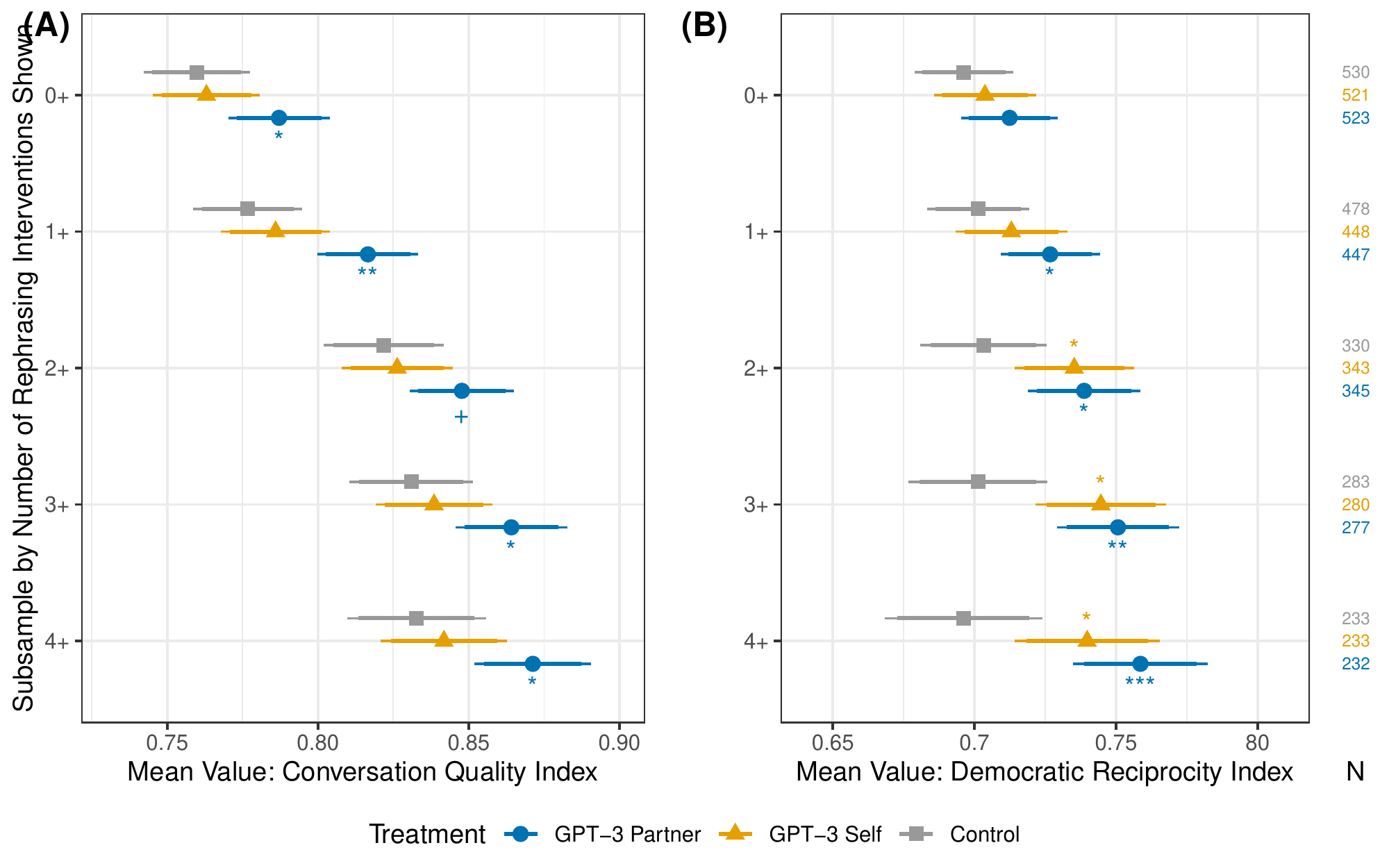}
\caption{\textbf{Conversational Quality (Panel A) and Democratic Reciprocity (Panel B) subgroup means and confidence intervals.} Higher values signify greater quality/support. The number of rephrasing interventions are overlapping sets, such that 0+ includes all observations. 90\%, and 95\% confidence intervals are based on unadjusted standard errors. Significance indicators stem from t-test comparisons to the control condition in each subgroup: $+p<.1, *p<.05, **p<.01, ***p<.001$}
\label{fig:mainresults}
\end{figure*}

Figure \ref{fig:mainresults} presents rephrasing results, showing means and confidence intervals for both conversational quality (A) and democratic reciprocity (B) across five different subgroups. The first group, 0+, estimates the treatment effect based on random assignment for all those who entered the chat platform, regardless of the length of conversation and including participants who had a full-length conversation with those who did not have a long enough chat to receive any treatment intervention. Each subsequent estimate uses a smaller subgroup of individuals who had a longer conversation, enough to receive various dosages of the treatment: 1 or more rephrasing suggestions, 2+, 3+, or 4+ (full treatment). The N for each subgroup is listed on the far right side of the figure.

These results provide striking evidence for the impact of the AI-rephrasing treatment on both conversational quality and democratic reciprocity, particularly for the \textit{partners} of those who received the suggestions. For both outcomes, partners of individuals who received one or more rephrasing reported significantly higher conversation quality and willingness to grant their political opponents democratic reciprocity: a statistically significant increase of 4 percentage points in conversational quality and 2.5 percentage points in democratic reciprocity. The effect on democratic reciprocity grows for partners of those who receive more of the treatment, with effect sizes of roughly 6 percentage points for full treatment. Although not as causally-identified as the foregoing results, evidence in Section 8 of the Supporting Information shows that these treatment effects are largest for individuals in conversations that start with the most initial disagreement on gun policy, an encouraging finding that underlies the strength of the AI chat treatment.

The effects are weaker for those who actually received the rephrasing suggestions themselves, as indicated in the ``GPT-Self'' estimates. For conversation quality, this is not surprising, given that the nature of the intervention was to provide suggestions that promote the other person's perception of being understood, and therefore the respondents who received the suggestions may not have been as directly validated by their partner. Notably, in spite of not reporting a higher quality conversation, these individuals do report higher democratic reciprocity than individuals in the control group, at substantive levels near those of their partners (and statistically indistinguishable from them). As we discuss further in the conclusion, these results, when taken together, suggest important initial promise for the use of this intervention in strengthening commitment to a fundamental democratic norm of reciprocity. These effect sizes are comparable to human-intervention studies designed to promote democratic reciprocity \cite{broockman2016durably,munger2017tweetment,santoro2022promise,broockmankalla2022}. Unlike human-moderator approaches, however, this treatment can be easily scaled to online settings and implemented broadly. 

We also examined the effect of the treatment condition on the level of substantive change in participants' attitudes towards gun regulation, presenting these results in the Supporting Information. While we find a small amount of average movement as a result of these conversations, consistent with our expectations we find no evidence that the AI assistant generated any more attitude change for either the treated person or their partner relative to change in control conversations. We see this lack of a policy effect as reassuring, suggesting that LMs can improve conversations without manipulating respondents to hold any particular perspective.

\section*{Conclusion}

Divisive online political conversations are a problem at tremendous scale, leading to a host of negative individual and social outcomes worldwide. We provide evidence that, when carefully deployed, cutting-edge AI tools can address these problems at that same scale. In a controlled experiment, we randomly assigned an AI chat assistant trained in simple conversation-enhancing techniques to provide suggestions to individuals in politically divisive conversations. Our results provide compelling evidence that this simple intervention, which can be applied across a variety of online chat contexts, has the power to increase the quality of a specific conversational exchange -- a social good in itself -- and also enhance commitment to democratic reciprocity. With respect to the latter, respondents display higher levels of a willingness to understand and allow the expression of opposing viewpoints in the political system  \textit{in general}, and not just in the context of their single conversation partner. This suggests the possibility of cascading consequences from these interactions into other political spheres and norms. Although there may also eventually be diminishing returns, these results suggest that more exposure to the intervention generates larger effects.

Importantly, we find these results while not impinging on human agency.  
At each AI intervention point, respondents were allowed to choose whether to send an alternative, keep their original text, or edit their message. In this way, the AI chat agent played a role similar to that which a trained human moderator might play in a moderated conversation, but with important advantages: the chat agent could intervene \textit{before} treated participants sent their texts, with real-time suggestions specifically tailored to the context of their conversation.

Although we find treatment effects for both those assigned the chat assistant and their partners, these effects are strongest and most consistent for the \textit{partner}. This difference is due in large part to nature of the treatment itself, as the particular rephrasing styles were all targeted at helping one's partner in the conversation feel more understood and respected. Our field experiment design does not allow us to explore additional reasons for this difference, a task that should be pursued in future research. 

Though many are rightly worried about the prospect of artificial intelligence being used to spread misinformation or polarize online communications, our findings indicate it may also be useful for promoting respect, understanding, and democratic reciprocity. We encourage future research into the ways that advances in technological tools like LMs can be used to address (rather than just exacerbate) political conflicts and crises facing democratic societies across the globe.

\matmethods{We recruited a nationally diverse sample via the survey firm Bovitz. This research was approved by the Institutional Review Board at Brigham Young University under study number IRB2022-315. All participants provided informed consent prior to participation. Participants first completed a short pre-survey, which ended with this common measure of feelings about gun policy in the United States: ``Which of the following statements comes closest to your overall view of gun laws in the United States? Gun laws should be MORE strict than they are today \textbackslash Gun laws are about right \textbackslash Gun laws should be LESS strict than they are today.'' 

Participants were then routed to our custom-built online chat platform where those who indicated gun laws should be ``more strict'' were paired for conversation with participants who said gun laws were ``about right'' or should be ``less strict.'' We combined these later respondents into one group for purposes of conversation matching based on responses to this question at Pew \cite{pew2021} and  in other surveys we ran in 2022 that included this question. These surveys suggested that these two groups together comprise roughly half of Americans, with the other half supporting more strict gun laws. We did not pair individuals who agreed on gun control in any conversations.

In the treatment condition, one (and only one) partner assigned to receive suggestions from the LM was shown a brief tutorial to orient them to the rephrasing process. During the conversation, treated partners received a rephrasing prompt for the first message longer than four words in every other conversation turn, regardless of the specific tone or content of the message. The rephrasing window provided participants three suggested alternatives (validating, restating, politeness -- in random order) to what they wrote.

In the post-conversation survey, respondents provided evaluations of the conversation quality, democratic reciprocity, and their positions on gun control policy. Our measure of conversation quality differs from how this concept is often measured in the deliberation literature \cite{steenbergen2003measuring,knobloch2022deliberative}, where conversations are often held face-to-face in groups (rather than by text communication, anonymously, and in dyads, as in our experiment) and where quality is measured by recording the number of interruptions in the group's conversation, the rigor of arguments and evidence, whether the group reached a mutually-agreed upon solution, and so forth. Our conception of conversation quality is instead much more similar to that proposed by Yeomans et.~al.~\cite{YEOMANS2020}. A full list of these items can be found in the Supporting Information, where we also provide analysis results for each item separately. 

Nearly 3 months after the experiment, we recontacted participants to explore the durability of these treatment effects. Consistent with the broader literature on the effects of brief contact interventions \cite{paluck2021} suggesting the need for ongoing intervention to generate ongoing effects, we found no evidence for persistence of these effects (see the Supporting Information).

Replication data and code files can be found at: https://osf.io/63zg2/. 

}

\showmatmethods{} 

\acknow{Funds for this research were provided by the National Science Foundation (award number 2141680), Duke University, and Brigham Young University.}

\showacknow{}
\newpage

\textbf{References}
\bibliography{pnas-sample}

\begin{thebibliography}{10}

\bibitem{doi:10.1177/2057047316628310}
DV Shah, Conversation is the soul of democracy: Expression effects, communication mediation, and digital media.
\newblock {\em\protect\JournalTitle{Communication and the Public}} \textbf{1}, 12--18 (2016).

\bibitem{deweypublicproblems}
J Dewey, {\em The public and its problems: An essay in political inquiry}.
\newblock (Penn State University Press), (2012).

\bibitem{muddiman2017personal}
A Muddiman, Personal and public levels of political incivility.
\newblock {\em\protect\JournalTitle{International Journal of Communication}} \textbf{11}, 21 (2017).

\bibitem{gutmann2004deliberative}
A Gutmann, D Thompson, {\em Why deliberative democracy?}
\newblock (Princeton University Press), (2004).

\bibitem{sydnor2019disrespectful}
E Sydnor, {\em Disrespectful democracy: The psychology of political incivility}.
\newblock (Columbia University Press), (2019).

\bibitem{finkel_political_2020}
EJ Finkel, et~al., {Political sectarianism in America}.
\newblock {\em\protect\JournalTitle{Science}} \textbf{370}, 533--536 (2020).

\bibitem{mutz2002cross}
DC Mutz, Cross-cutting social networks: Testing democratic theory in practice.
\newblock {\em\protect\JournalTitle{American Political Science Review}} \textbf{96}, 111--126 (2002).

\bibitem{jacobs2009talking}
LR Jacobs, FL Cook, MXD Carpini, {\em Talking together: Public deliberation and political participation in America}.
\newblock (University of Chicago Press), (2009).

\bibitem{mosseri2020}
A Mosseri, Say hi to messenger: Introducing new messaging features for instagram.
\newblock {\em\protect\JournalTitle{Meta Newsroom}} (2020).

\bibitem{bleu2023}
N Bleu, 27 latest {F}acebook {M}essenger statistics (2023 edition).
\newblock {\em\protect\JournalTitle{BloggingWizard}} (2023).

\bibitem{harlow2011}
S Harlow, Social media and social movements: Facebook and an online guatemalan justice movement that moved offline.
\newblock {\em\protect\JournalTitle{New Media and Society}} \textbf{14}, 225--243 (2011).

\bibitem{mundt2018}
M Mundt, K Ross, CM Burnett, Scaling social movements through social media: The case of black lives matter.
\newblock {\em\protect\JournalTitle{Social Media + Society}} \textbf{October}, 1--14 (2018).

\bibitem{jiao2020}
P Jiao, A Veiga, A Walther, Social media, news media and the stock market.
\newblock {\em\protect\JournalTitle{Journal of Economic Behavior and Organization}} \textbf{176}, 63--90 (2020).

\bibitem{kessel2020}
P van Kessel, R Widjaya, S Shah, A Smith, A Hughes, Congress soars to new heights on social media.
\newblock {\em\protect\JournalTitle{Pew Research Center}} (2020).

\bibitem{garsd2019}
J Garsd, In an increasingly polarized {A}merica, is it possible to be civil on social media?
\newblock {\em\protect\JournalTitle{National Public Radio}} (2019).

\bibitem{settle2018}
JE Settle, {\em Frenemies: How social media polarizes America}.
\newblock (Cambridge University Press), (2018).

\bibitem{bail2021}
C Bail, {\em Breaking the social media prism: How to make our platforms less polarizing}.
\newblock (Princeton University Press), (2021).

\bibitem{sun2021}
Q Sun, M Wojcieszak, S Davidson, Over-time trends in incivility on social media: Evidence from political, non-political, and mixed sub-reddits over eleven years.
\newblock {\em\protect\JournalTitle{Frontiers in Political Science}} \textbf{3}, 741605 (2021).

\bibitem{rainie2012}
L Rainie, A Lenhart, A Smith, The tone of life on social networking sites.
\newblock {\em\protect\JournalTitle{Pew Research Center}} (2012).

\bibitem{millerconover2015}
PR Miller, PJ Conover, Why partisan warriors don't listen: The gendered dynamics of intergroup anxiety and partisan conflict.
\newblock {\em\protect\JournalTitle{Politics, Groups, and Identities}} \textbf{3}, 21--39 (2015).

\bibitem{lee2021}
AHY Lee, How the politicization of everyday activities affects the public sphere: The effects of partisan stereotypes on cross-cutting interactions.
\newblock {\em\protect\JournalTitle{Political Communication}} \textbf{38}, 499--518 (2021).

\bibitem{santoro2022promise}
E Santoro, DE Broockman, The promise and pitfalls of cross-partisan conversations for reducing affective polarization: Evidence from randomized experiments.
\newblock {\em\protect\JournalTitle{Science Advances}} \textbf{8}, eabn5515 (2022).

\bibitem{carlson2022goes}
TN Carlson, JE Settle, {\em What goes without saying}.
\newblock (Cambridge University Press), (2022).

\bibitem{Kalmoe2018}
NP Kalmoe, JR Gubler, DA Wood, Toward conflict or compromise? {H}ow violent metaphors polarize partisan issue attitudes.
\newblock {\em\protect\JournalTitle{Political Communication}} \textbf{35}, 333--352 (2018).

\bibitem{kalmoe2022radical}
NP Kalmoe, L Mason, {\em Radical American partisanship: Mapping violent hostility, its causes, and the consequences for democracy}.
\newblock (University of Chicago Press), (2022).

\bibitem{krupnikov2022other}
Y Krupnikov, JB Ryan, {\em The other divide: Polarization and disengagement in American politics}.
\newblock (Cambridge University Press), (2022).

\bibitem{wolf2012incivility}
MR Wolf, JC Strachan, DM Shea, Incivility and standing firm: A second layer of partisan division.
\newblock {\em\protect\JournalTitle{PS: Political Science \& Politics}} \textbf{45}, 428--434 (2012).

\bibitem{mendelberg_2002}
T Mendelberg, The deliberative citizen: {Theory} and evidence in {\em Political decision making, deliberation and participation}, eds.{} MX Delli~Carpini, L Huddy, RY Shapiro.
\newblock (Jawe Press, Greenwich, CT) Vol.{}~6, pp. 151--193 (2002).

\bibitem{scudder2020beyond}
MF Scudder, {\em Beyond empathy and inclusion: The challenge of listening in democratic deliberation}.
\newblock (Oxford University Press), (2020).

\bibitem{dobson2014listening}
A Dobson, {\em Listening for democracy : Recognition, representation, reconciliation}.
\newblock (Oxford University Press), (2014).

\bibitem{broockman2016durably}
D Broockman, J Kalla, Durably reducing transphobia: A field experiment on door-to-door canvassing.
\newblock {\em\protect\JournalTitle{Science}} \textbf{352}, 220--224 (2016).

\bibitem{levenduskystecula2021}
MS Levendusky, DA Stecula, {\em We need to talk}.
\newblock (Cambridge University Press), (2022).

\bibitem{wojcieszak2020}
M Wojcieszak, BR Warner, Can interparty contact reduce affective polarization? {A} systematic test of different forms of intergroup contact.
\newblock {\em\protect\JournalTitle{Political Communication}} \textbf{37}, 789--811 (2020).

\bibitem{amsalem2022}
E Amsalem, E Merkley, PJ Loewen, Does talking to the other side reduce inter-party hostility? {E}vidence from three studies.
\newblock {\em\protect\JournalTitle{Political Communication}} \textbf{39}, 61--78 (2022).

\bibitem{reis2017toward}
HT Reis, EP Lemay~Jr, C Finkenauer, Toward understanding understanding: The importance of feeling understood in relationships.
\newblock {\em\protect\JournalTitle{Social and Personality Psychology Compass}} \textbf{11}, e12308 (2017).

\bibitem{ruan2020can}
Y Ruan, HT Reis, MS Clark, JL Hirsch, BD Bink, Can i tell you how i feel? {P}erceived partner responsiveness encourages emotional expression.
\newblock {\em\protect\JournalTitle{Emotion}} \textbf{20}, 329 (2020).

\bibitem{itzchakov2022foster}
G Itzchakov, HT Reis, N Weinstein, How to foster perceived partner responsiveness: High-quality listening is key.
\newblock {\em\protect\JournalTitle{Social and Personality Psychology Compass}} \textbf{16}, e12648 (2022).

\bibitem{livingstone2020they}
AG Livingstone, L Fern{\'a}ndez~Rodr{\'\i}guez, A Rothers, “{T}hey just don't understand us”: The role of felt understanding in intergroup relations.
\newblock {\em\protect\JournalTitle{Journal of personality and social psychology}} \textbf{119}, 633 (2020).

\bibitem{mutz2006hearing}
DC Mutz, {\em Hearing the other side: Deliberative versus participatory democracy}.
\newblock (Cambridge University Press), (2006).

\bibitem{tappin_wittenberg_hewitt_berinsky_rand_2022}
BM Tappin, C Wittenberg, L Hewitt, a berinsky, DG Rand, Quantifying the persuasive returns to political microtargeting (2022).

\bibitem{aggarwal20232}
M Aggarwal, et~al., A 2 million-person, campaign-wide field experiment shows how digital advertising affects voter turnout.
\newblock {\em\protect\JournalTitle{Nature Human Behaviour}} pp. 1--10 (2023).

\bibitem{munger2017tweetment}
K Munger, Tweetment effects on the tweeted: Experimentally reducing racist harassment.
\newblock {\em\protect\JournalTitle{Political Behavior}} \textbf{39}, 629--649 (2017).

\bibitem{bail2018exposure}
CA Bail, et~al., Exposure to opposing views on social media can increase political polarization.
\newblock {\em\protect\JournalTitle{Proceedings of the National Academy of Sciences}} \textbf{115}, 9216--9221 (2018).

\bibitem{tiku2022}
N Tiku, GD Vynck, W Oremus, Big tech was moving cautiously on {AI}. {T}hen came {C}hat{GPT}.
\newblock {\em\protect\JournalTitle{The Washington Post}} (2022).

\bibitem{argyle22}
LP Argyle, et~al., Out of one, many: Using language models to simulate human samples.
\newblock {\em\protect\JournalTitle{Political Analysis}} \textbf{31} (2023).

\bibitem{brown2020language}
TB Brown, et~al., Language models are few-shot learners.
\newblock {\em\protect\JournalTitle{arXiv:2005.14165}} (2020).

\bibitem{bender2021dangers}
EM Bender, T Gebru, A McMillan-Major, S Shmitchell, On the dangers of stochastic parrots: Can language models be too big? in {\em Proceedings of the 2021 ACM conference on fairness, accountability, and transparency}.
\newblock pp. 610--623 (2021).

\bibitem{kleinberg_human_2018}
J Kleinberg, H Lakkaraju, J Leskovec, J Ludwig, S Mullainathan, Human decisions and machine predictions.
\newblock {\em\protect\JournalTitle{The Quarterly Journal of Economics}} \textbf{133}, 237--293 (2018).

\bibitem{panchArtificialIntelligenceAlgorithmic2019}
T Panch, H Mattie, R Atun, Artificial intelligence and algorithmic bias: Implications for health systems.
\newblock {\em\protect\JournalTitle{Journal of Global Health}} \textbf{9}, 010318 (2019).

\bibitem{caliskan2017semantics}
A Caliskan, JJ Bryson, A Narayanan, Semantics derived automatically from language corpora contain human-like biases.
\newblock {\em\protect\JournalTitle{Science}} \textbf{356}, 183--186 (2017).

\bibitem{obermeyerDissectingRacialBias2019}
Z Obermeyer, B Powers, C Vogeli, S Mullainathan, Dissecting racial bias in an algorithm used to manage the health of populations.
\newblock {\em\protect\JournalTitle{Science}} \textbf{366}, 447--453 (2019).

\bibitem{sharma2023human}
A Sharma, IW Lin, AS Miner, DC Atkins, T Althoff, Human--{AI} collaboration enables more empathic conversations in text-based peer-to-peer mental health support.
\newblock {\em\protect\JournalTitle{Nature Machine Intelligence}} pp. 1--12 (2023).

\bibitem{kriplean2012}
T Kriplean, M Toomim, J Morgan, A Borning, AJ Ko, Is this what you meant?: Promoting listening on the web with reflect.
\newblock {\em\protect\JournalTitle{CHI '12: Proceedings of the SIGCHI Conference on Human Factors in Computing Systems}} pp. 1559--1568 (2012).

\bibitem{kim2021}
S Kim, J Eun, J Serring, J Lee, Moderator chatbot for deliberative discussion: Effects of discussion structure and discussant facilitation.
\newblock {\em\protect\JournalTitle{Proceedings of ACM Human-Computer Interactions}} \textbf{5} (2021).

\bibitem{YEOMANS2020}
M Yeomans, J Minson, H Collins, F Chen, F Gino, Conversational receptiveness: Improving engagement with opposing views.
\newblock {\em\protect\JournalTitle{Organizational Behavior and Human Decision Processes}} \textbf{160}, 131--148 (2020).

\bibitem{fishkin2019deliberative}
J Fishkin, et~al., Deliberative democracy with the online deliberation platform in {\em The 7th AAAI Conference on Human Computation and Crowdsourcing (HCOMP 2019).}
\newblock (2019).

\bibitem{esaiasson2017responsiveness}
P Esaissaon, M Gilljam, M Persson, Responsiveness beyond policy satisfaction: Does it matter to citizens?
\newblock {\em\protect\JournalTitle{Comparative Political Studies}} \textbf{50}, 739--765 (2017).

\bibitem{paluck2019contact}
EL Paluck, SA Green, DP Green, The contact hypothesis re-evaluated.
\newblock {\em\protect\JournalTitle{Behavioural Public Policy}} \textbf{3}, 129--158 (2019).

\bibitem{gordon2016}
AM Gordon, S Chen, Do you get where i’m coming from?: Perceived understanding buffers against the negative impact of conflict on relationship satisfaction.
\newblock {\em\protect\JournalTitle{Journal of Personality and Social Psychology}} \textbf{110}, 239 (2016).

\bibitem{pollmann2009}
MMH Pollmann, C Fikenauer, Investigating the role of two types of understanding in relationship well-being: Understanding is more important than knowledge.
\newblock {\em\protect\JournalTitle{Personality and Social Psychology Bulletin}} \textbf{35}, 1512--1527 (2009).

\bibitem{minson2022}
JA Minson, FS Chen, Receptiveness to opposing views: Conceptualization and integrative review.
\newblock {\em\protect\JournalTitle{Personality and Social Psychology Review}} \textbf{26}, 93--111 (2022).

\bibitem{hartman2022interventions}
R Hartman, et~al., Interventions to reduce partisan animosity.
\newblock {\em\protect\JournalTitle{Nature human behaviour}} \textbf{6}, 1194--1205 (2022).

\bibitem{broockmankalla2022}
D Broockman, J Kalla, S Westwood, Does affective polarization undermine democratic norms or accountability? {M}aybe not.
\newblock {\em\protect\JournalTitle{American Journal of Political Science}} (2022).

\bibitem{kreiss2023polarization}
D Kreiss, SC McGregor, A review and provocation: On polarization and platforms.
\newblock {\em\protect\JournalTitle{New Media \& Society}} (2023).

\bibitem{obrien2013}
K O'Brien, W Forrest, D Lynott, M Daly, Racism, gun ownership and gun control: Biased attitudes in us whites may influence policy decisions.
\newblock {\em\protect\JournalTitle{PLOS ONE}} \textbf{8}, e77552 (2013).

\bibitem{lacombe2019}
MJ Lacombe, AJ Howat, JE Rothschild, Gun ownership as a social identity: Estimating behavioral and attitudinal relationships.
\newblock {\em\protect\JournalTitle{Social Science Quarterly}} \textbf{100}, 2408--2424 (2019).

\bibitem{filindra2021}
A Filindra, NJ Kaplan, BE Buyuker, Racial resentment or sexism? {W}hite {A}mericans’ outgroup attitudes as predictors of gun ownership and nra membership.
\newblock {\em\protect\JournalTitle{Sociological inquiry}} \textbf{91}, 253--286 (2021).

\bibitem{lacombe2021}
MJ Lacombe, {\em Firepower: How the NRA turned gun owners into a political force}.
\newblock (Princeton University Press), (2021).

\bibitem{yeomans2018}
M Yeomans, A Kantor, D Tingley, The politeness package: Detecting politeness in natural language.
\newblock {\em\protect\JournalTitle{The R Journal}} \textbf{10}, 489--502 (2008).

\bibitem{gerber2012field}
AS Gerber, DP Green, {\em Field experiments: Design, analysis, and interpretation}.
\newblock (WW Norton), (2012).

\bibitem{pew2021}
PR Center, Amid a series of mass shootings in the {U.S.}, gun policy remains deeply divisive.
\newblock {\em\protect\JournalTitle{Pew Research Center}} (2021).

\bibitem{steenbergen2003measuring}
MR Steenbergen, A B{\"a}chtiger, M Sp{\"o}rndli, J Steiner, Measuring political deliberation: A discourse quality index.
\newblock {\em\protect\JournalTitle{Comparative European Politics}} \textbf{1}, 21--48 (2003).

\bibitem{knobloch2022deliberative}
K Knobloch, J Gastil, KR Knobloch, How deliberative experiences shape subjective outcomes: A study of fifteen minipublics from 2010-2018.
\newblock {\em\protect\JournalTitle{Journal of Deliberative Democracy}} \textbf{18} (2022).

\bibitem{paluck2021}
EL Paluck, R Porat, CS Clark, DP Green, Prejudice reduction: Progress and challenges.
\newblock {\em\protect\JournalTitle{Annual Review of Psychology}} \textbf{72}, 533--560 (2021).

\end{thebibliography}

\end{document}